\documentclass[aps,pre,floats,twocolumn,showpacs,superscriptaddress]{revtex4}
\usepackage{graphicx}
\usepackage{rotate}
\begin{document}

\title{Effect of size heterogeneity on community identification in complex networks}

\author{Leon Danon}

\affiliation {Departament de F\'{i}sica Fonamental,Universitat de
Barcelona, Marti i Franques 1, 08028 Barcelona, Spain}

\author{Albert D\'{i}az-Guilera}

\affiliation {Departament de F\'{i}sica Fonamental,Universitat de
Barcelona, Marti i Franques 1, 08028 Barcelona, Spain}

\author{Alex Arenas}

\affiliation{Departament d'Enginyeria Inform\`{a}tica i
Matem\`{a}tiques, Campus Sescelades, Universitat Rovira i Virgili,
43007 Tarragona, Spain}


\begin{abstract}
Identifying community structure can be a potent tool in the analysis
and understanding of the structure of complex networks. Up to now,
methods for evaluating the performance of identification algorithms
use ad-hoc networks with communities of equal size. We show that
inhomogeneities in community sizes can and do affect the performance
of algorithms considerably, and propose an alternative method which
takes these factors into account. Furthermore, we propose a simple
modification of the algorithm proposed by Newman for community
detection (Phys. Rev. E {\bf 69} 066133) which treats communities of
different sizes on an equal footing, and show that it outperforms
the original algorithm while retaining its speed.

\end{abstract}

\maketitle
\section{Introduction}

Natural and artificial systems often have architectures which are
best described as complex networks. The topologies of networks have
been extensively studied in various disciplines in recent years,
particularly within physics
\cite{BARev,DMRev,NRev,Strogatz01,Boccaletti06}. A part of that
research has been directed at the study of modules or communities in
networks. Communities can be defined as subsets of nodes which are
densely connected to each other and loosely connected to the rest of
the network. Such structures have been discovered in networks as
diverse as banking networks, metabolic networks, the airport network
and most notably in social networks \cite{Thurner04,
Ravasz02,Guimera05a,Guimera05b,GN}.

Despite efforts spanning several decades in this direction
\cite{KernighanLin,PothenRev}, the identification of community
structure in networks remains an open problem. The space of possible
partitions of even a small network is very large indeed. Several
methods have been proposed for finding meaningful partitions in
networks of reasonable size. These methods vary considerably from
one another, not only in their general approach, but also in
sensitivity and computational effort (for recent reviews, see
\cite{Newman04,Danon05} and chapter 7.1 of \cite{Boccaletti06}). In
general, those methods which are more accurate tend to be able to
explore a larger portion of the partition space, and are therefore
computationally expensive (see for example
\cite{Reichardt04,Guimera04}). On the other hand, those methods
which explore a smaller region of the partition space tend to be
faster, but as a consequence, less accurate
\cite{Newman04a,Clauset04}. The challenge, therefore, is to find
methods which are both fast and accurate, and several attempts have
been made \cite{Duch05,Wu03,Bagrow04}.

In this paper we reevaluate the benchmark most commonly used at
present to measure the sensitivity of a particular community
identification algorithm \cite{NG}. This benchmark, although useful,
does not take into account the fact that networks exhibit community
structure where the community sizes are highly skewed, despite the
fact that several authors have observed that distributions of
community sizes seem to follow power laws in many cases
\cite{Guimera03b,Gleiser03,Arenas03,Newman04a,Clauset04,Palla05}. In
the next section we propose a benchmark for measuring algorithm
sensitivity which takes this skew into account. In section
\ref{sec:modfast} we examine Newman's Fast algorithm (NF) for
community detection \cite{Newman04a}, and see that it is affected by
a skew in the community size distribution, showing a tendency to
find large communities at the expense of smaller ones. We propose a
modification of the algorithm, in which the communities of different
sizes are treated equally, and in section \ref{sec:results} we show
that it outperforms the NF algorithm in sensitivity, with no
tradeoff in terms of computational effort.

\section{Evaluating algorithm performance on ad-hoc networks}
\label{sec:bench}

To quantify how good a particular network partition is, the
modularity measure $Q$ was introduced in \cite{NG}, and has been
widely used since then. Based on a predefined set of communities $i$
in a network, a community connection matrix $e_{ij}$ is defined,
where each member represents the proportion of links from community
$i$ to community $j$. Note that the matrix is normalised, that is,
each of the members of the matrix $e_{ij}=\frac{L_{ij}}{L_{total}}$,
$L_{ij}$ being the number of links between community $i$ and
community $j$, and $L_{total}$ is the total number of links in the
network \cite{NG}. The proportion of links belonging to community
$i$ is denoted $a_i$ and is simply the sum, $a_i=\sum_ie_{ij}$. The
computation of $Q$ is as follows:

\begin{equation}
  Q=\sum_j (e_{ii} - a_i^2)
\end{equation}

The modularity, $Q$, quantifies the difference between the
intra-community links and the expected value for the same
communities in a randomised network. Note that the modularity is a
relative value, and while it gives an idea of how good a partition
of the network is, it cannot tell us whether this partition is the
best one possible. It does provide a useful way of comparing the
performance of different community identification algorithms applied
on one particular network.

The method most commonly used to compare the sensitivity of
community identification methods was also proposed in \cite{NG}, and
is independent of the modularity measure. It uses a benchmark test
based on networks typically containing 128 nodes grouped into four
communities which contain the same number of nodes, 32, and links
(on average 16 per node, $k = 16$). Pairs of nodes belonging to the
same community are linked with probability $p_{in}$, whereas pairs
belonging to different communities are joined with probability
$p_{out}$. The value of $p_{out}$ controls the average number of
links a node has to members of any other community, $z_{out}$. While
$p_{out}$ (and therefore $z_{out}$) is varied freely, the value of
$p_{in}$ is chosen to keep the total average node degree $k$
constant. As $z_{out}$ is increased from zero, the communities
become more and more fuzzy and harder to identify. Different
community detection algorithms, when applied to these networks may
give different results, reflecting their sensitivity. Since the
`real' community structure is well known in this case, it is
possible to measure how well the partitions the algorithm finds
compare to the original partitions.

Here we use a measure based on information theory for this purpose.
The normalised mutual information, $I(A,B)$, explicitly measures the
amount of information about partition A that is gained by knowing
partition B \cite{Fred03,Kuncheva04}. In other words, it is the
amount of information the algorithm is able to extract from the
pre-defined partition just from the topology. \cite{Danon05}. This
independent measure is based on defining a {\it confusion matrix}
$\bf{M}$, where rows correspond to ``real'' communities, and columns
correspond to ``found'' communities. The element of $\bf{M}$,
$M_{ij}$ is the number of nodes in the real community $i$ that
appear in the found community $j$. A measure of similarity between
the partitions, is then:

\begin{equation}
I(A,B)=\frac{-2\sum^{c_A}_{i=1}\sum^{c_B}_{j=1}
M_{ij}\log\left(\frac{M_{ij}N}{M_{i.}M_{.j}}\right)}
{\sum^{c_A}_{i=1}M_{i.}\log\left(\frac{M_{i.}}{N}\right)
 + \sum^{c_B}_{j=1}M_{.j}\log\left(\frac{M_{.j}}{N}\right)}
\end{equation}

where the number of real communities is denoted $c_A$ and the number
of found communities is denoted $c_B$, $N$ is the number of nodes,
the sum over row $i$ of matrix $M_{ij}$ is denoted $M_{i.}$ and the
sum over column $j$ is denoted $M_{.j}$.

\begin{figure}
  \includegraphics*[width=\columnwidth]{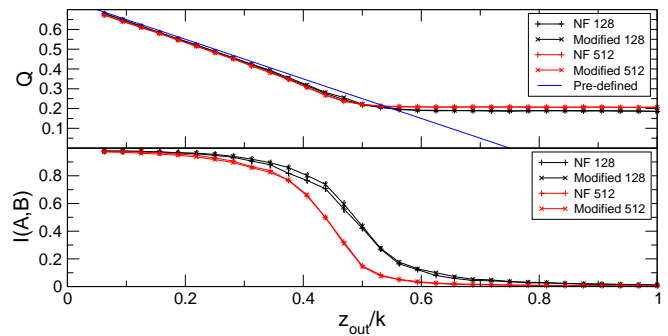}\\
  \caption{(Colour online) Sensitivity of the NF algorithm and the
  modification described in Section \ref{sec:modfast},
  applied to ad-hoc networks with four equal-sized
  communities, for two network sizes, 128 nodes and 512 nodes,
  with average degree $k=16$. The top figure shows the variation of modularity
  found by the algorithms with $z_{out}/k$. For low
  values of $z_{out}/k$, the value of $Q$ of the partitions found closely follow
  the expected modularity. For higher values of
  $z_{out}/k$, the partitions found show a better
  modularity than pre-defined partitions. There is little difference
  between results for different network sizes. In the bottom figure the
  comparison between pre-defined and found partitions using the mutual
  information measure $I(A,B)$ is shown. Both algorithms have similar
  sensitivity for both network sizes, but the sensitivity is reduced at the
  same value of $z_{out}/k$ for the larger network, suggesting that
  communities are more fuzzy the larger they are as discussed in the text.
  }\label{fig:4coms}
\end{figure}

Because of the particular definition of these ad-hoc networks, it is
tempting to think that similar networks with four communities
sharing the same value of $z_{out}/k$ will have an equivalent
community structure, and that a particular method of community
identification will perform equally well. This, however, is highly
dependent on the number of nodes that the network has, and more
importantly the number of nodes in each community. For example a
network with 128 nodes with four communities each of size 32 with
$k=16$ and $z_{out} = 6$, say, will have a better defined community
structure than a network with the same values of $k$ and $z_{out}$
which is comprised of 512 nodes with four communities each of size
128. This is simply due to the fact that the internal links are
spread out over a larger number of nodes, thus making the
communities less dense, in terms of proportion of actual links to
possible links. In Figure \ref{fig:4coms}, we can see that the same
algorithm will perform significantly better on a network with $128$
nodes than on one with 512 nodes with the same values of $k$ and
$z_{out}$.

Furthermore, in real networks the distribution of community sizes is
highly skewed, and has been observed to follow power laws in many
cases \cite{Guimera03b,Arenas03,Newman04,Clauset04,Palla05}. We
argue that this difference in sizes is important and affects
different identification algorithms in different ways. To be able to
evaluate the effect that a spread in community sizes will have on
the performance of any algorithm, we first need to be able to create
networks with controlled community structure of differing community
sizes.

Consider a set of $N_c$ communities where each community contains
$n_i$ nodes. Considering pairs of nodes, if both nodes are in the
same community a link is placed between them with probability
$P_{in}$, otherwise they are connected with probability $P_e$.
Should $P_{in}$ be constant for all communities, the number of links
of community $i$ would scale as the square of its size, $n_i^2$. To
give the same weight to communities of different sizes, we propose
that $P_{in}=F/n_i$ where $F$ is a control parameter. In this way we
are able to control both internal and external cohesion by varying
$F$ and $P_e$ respectively. This method of network creation is
equivalent to creating a random Erd\"{o}s- Renyi network with the
probability of linking being equal to $P_e$ and then superposing
$N_c$ random networks whose sizes correspond to $n_i$ where the
probability of internal linking is $F/n_i$.

\begin{figure}[h]
\centerline{
  \includegraphics*[width=0.4\columnwidth]{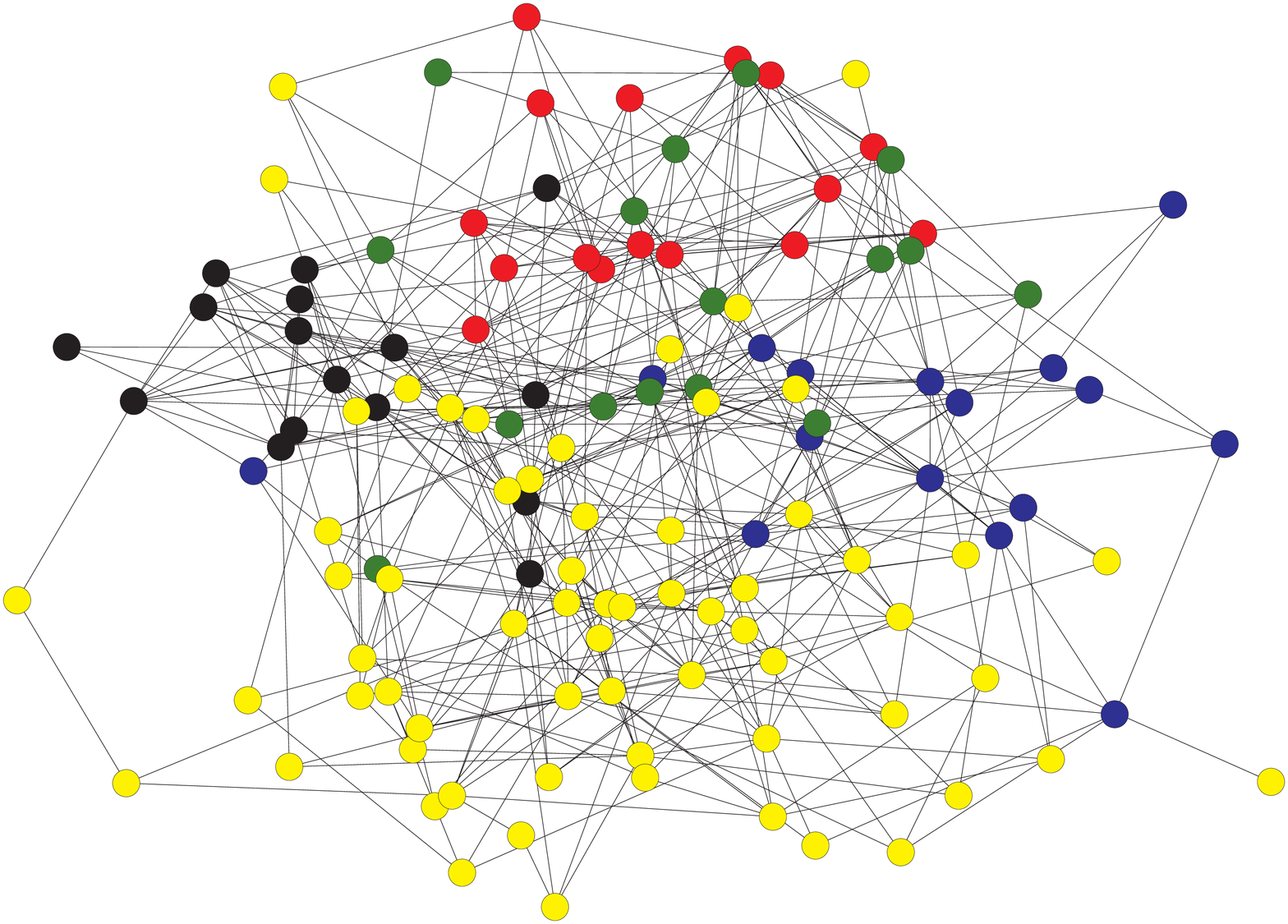}
  \includegraphics*[width=0.4\columnwidth]{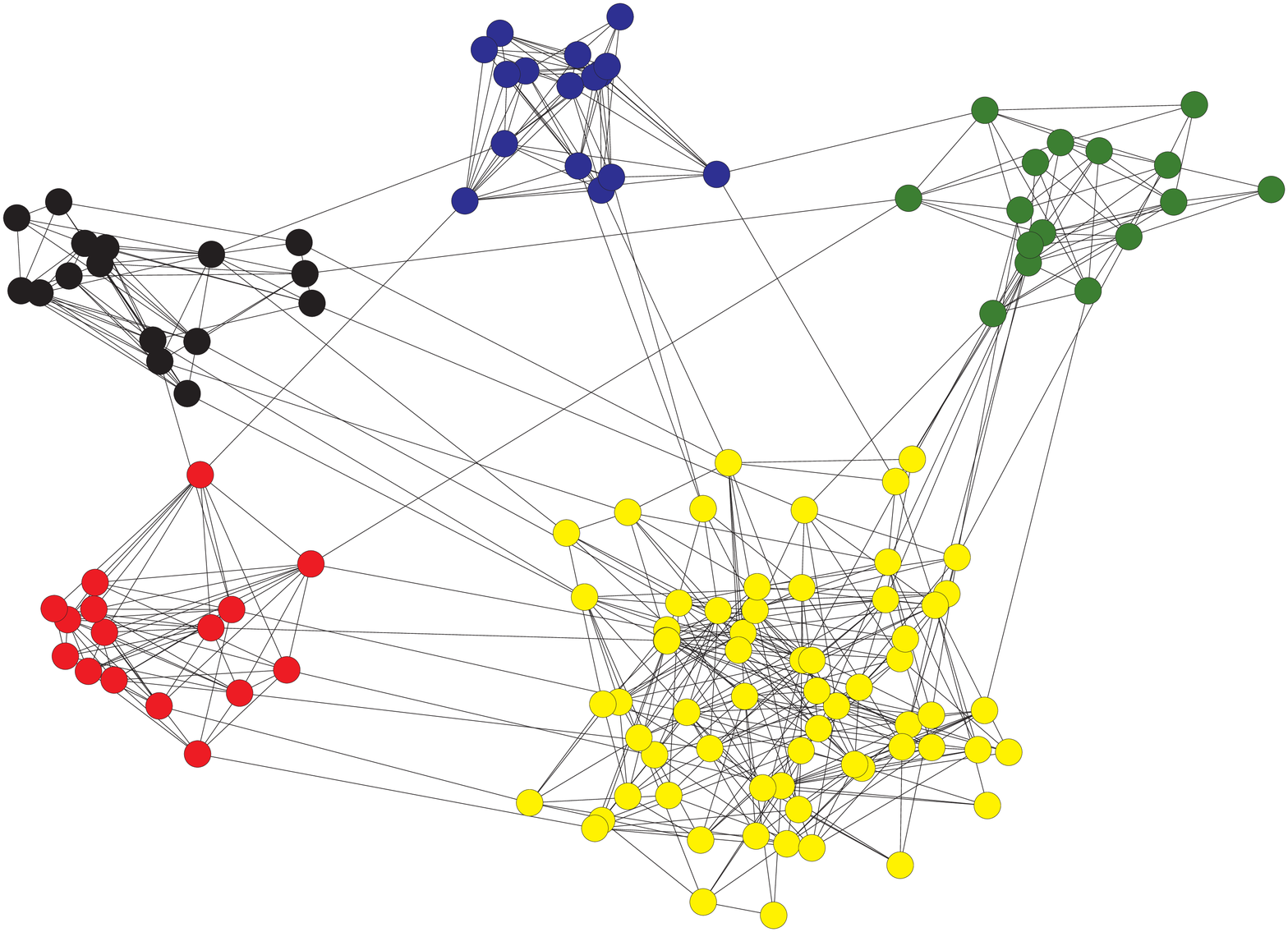}
} \centerline{(a)\hspace{0.4\columnwidth}(b)}
\centerline{\includegraphics*[width=0.9\columnwidth]{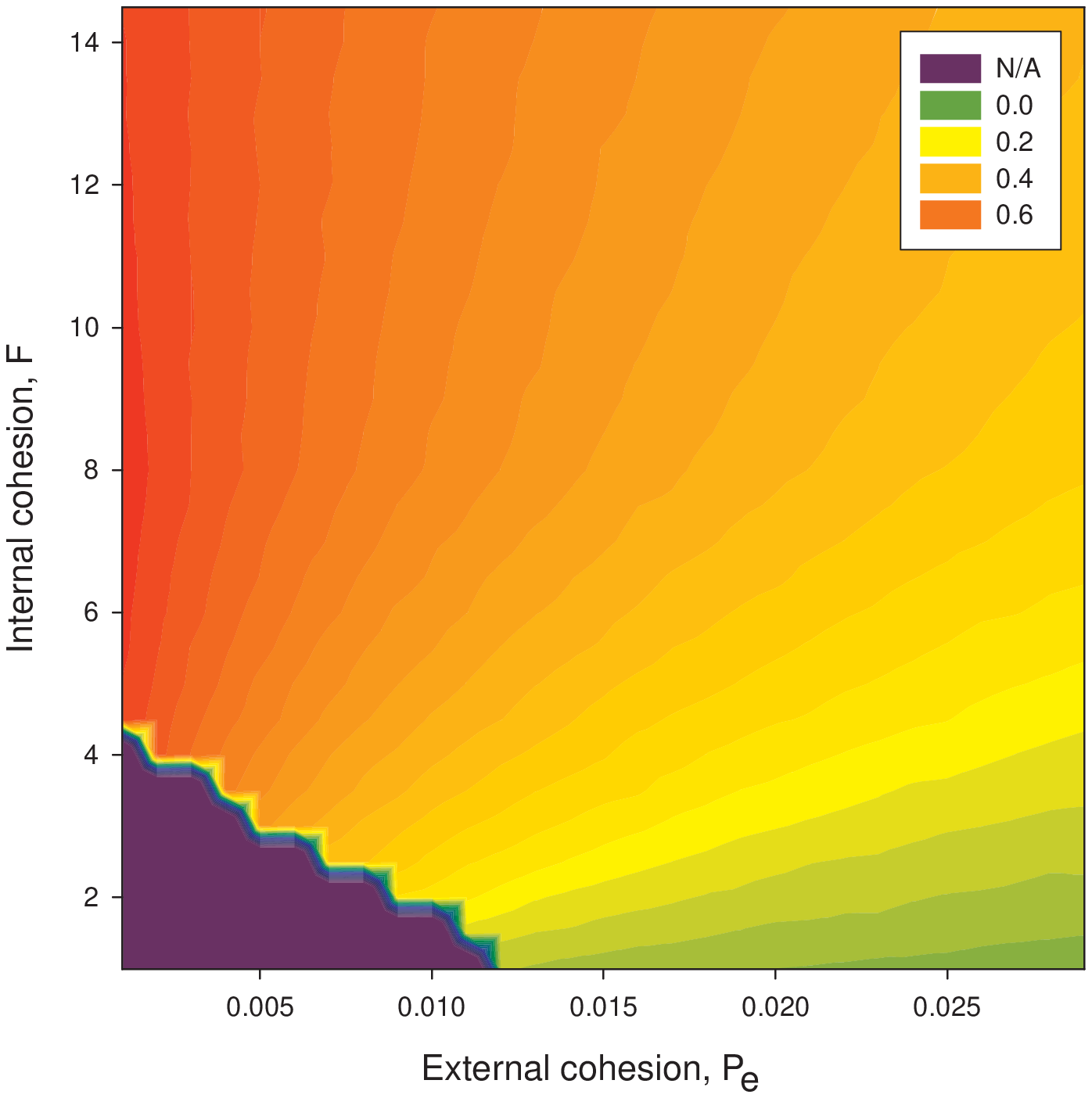}}
\centerline{(c)}
 \caption{
  (Colour online) Two examples networks created as described
  in the main text with 5 communities four of which have 16 nodes and
  one has 64, (a) has $P_e = 0.007$ and $F = 8$ and in (b) $P_e= 0.03$ and $F = 3$.
  (c) The modularity $Q$ of networks as generated in the main text for
  values of $P_e$ between 0.001 and 0.03, and values of $F$ between 1 and 14.
  The dark zones represent parts of the parameter space where the networks
  constructed were disconnected for more than 1 in 100 realisations.
\label{fig:adhocdiff}}
\end{figure}

Figure \ref{fig:adhocdiff}(a and b) shows two networks with 5
communities each, containing one community of 64 nodes and 4
communities of 16 nodes each for two different values of $P_e$ and
$F$. Figure \ref{fig:adhocdiff}c shows the value of $Q$ when the
network partition corresponds exactly to the prescribed communities
as a function of $F$ and $P_e$. While these community sizes are
chosen to be illustrative, this method of network creation is
completely general and community sizes can be drawn from any given
distribution.

\section{Dynamics of the Fast algorithm and its modification}
\label{sec:modfast}

The performance of various community identification algorithms has
recently been studied both in terms of speed and in terms of
accuracy. Having a method of generation of networks with communities
of differing sizes puts us in a position to test the way these sizes
can affect the performances of identification algorithms. In
particular we concentrate on Newman's Fast algorithm as proposed in
\cite{Newman04a}. It is dubbed fast since it runs in almost linear
time for sparse networks, $O(n\log^2{n})$ \cite{Clauset04}, and
while it is not the most accurate method, it remains the only
algorithm able to extract community structure information from very
large networks \cite{Danon05}.

Let us consider a network that has been partitioned in some
arbitrary way. Joining two neighbouring partitions $i$ and $j$,
would produce a change in modularity:

\begin{equation}
dQ_{ij}=2(e_{ij}-\frac{a_ia_j}{2L_{total}}) \label{dqpure}
\end{equation}

This can be interpreted as a measure of affinity of communities $i$
and $j$, and can subsequently be used to find the two communities
which are most alike (highest $dQ$). Starting with each node in the
network in its own community, one can join pairs of communities with
the highest $dQ$. This process can then be performed and repeated
until the whole network is contained in one community. As the author
states in \cite{Newman04}, this is very similar to agglomerative
hierarchical clustering methods \cite{Everitt93,Scott00}. Here,
``distance'' measures such as single linkage or complete linkage are
replaced by $dQ$. It also differs from hierarchical clustering in
that not all pairs of clusters are compared, only those connected by
real links in the network.

\begin{figure}[h]
\centerline{
\includegraphics*[width=0.4\columnwidth]{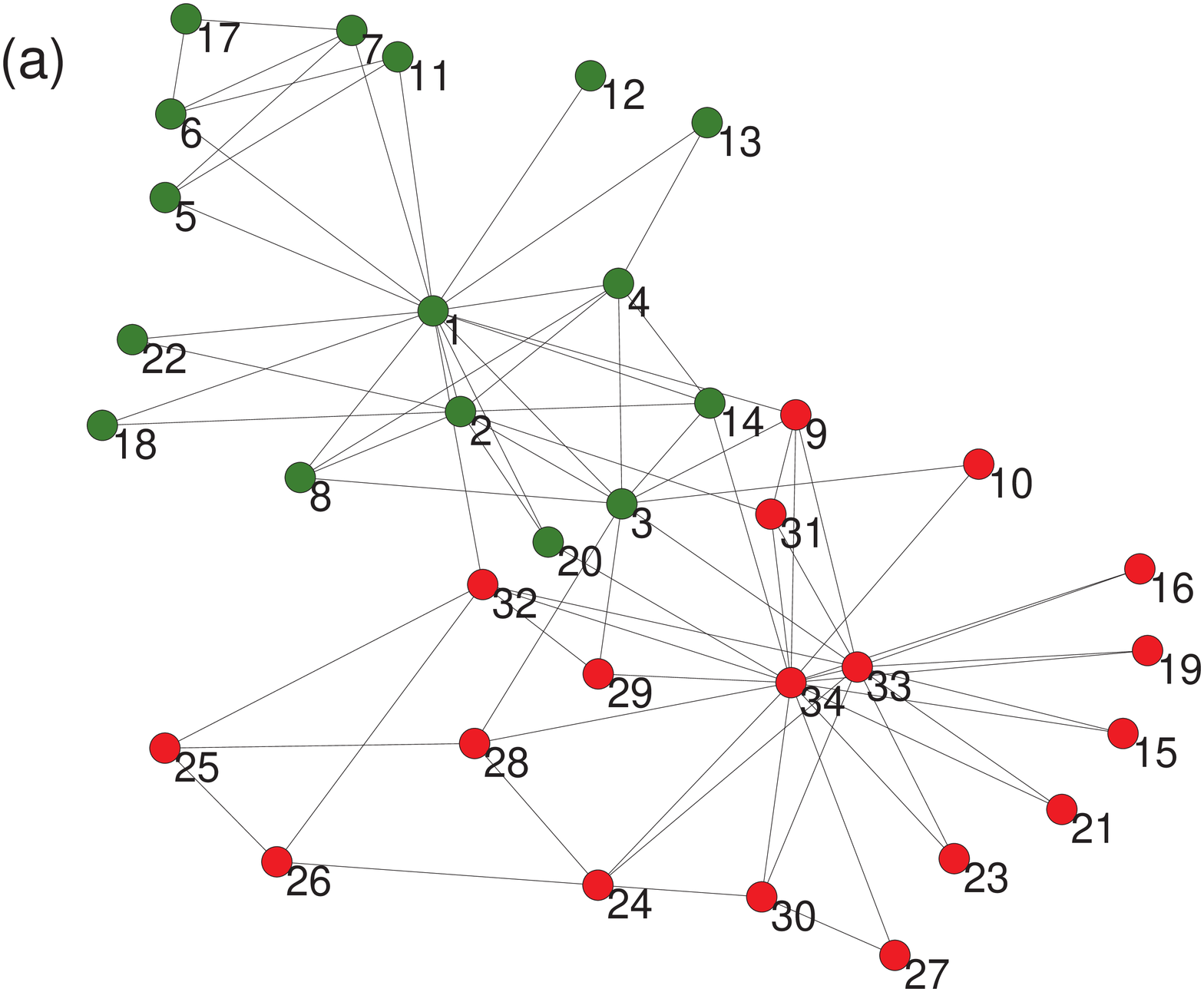}\hspace{0.12\columnwidth}
\includegraphics*[width=0.35\columnwidth]{fig3b.eps}
} \centerline{\includegraphics*[width=\columnwidth]{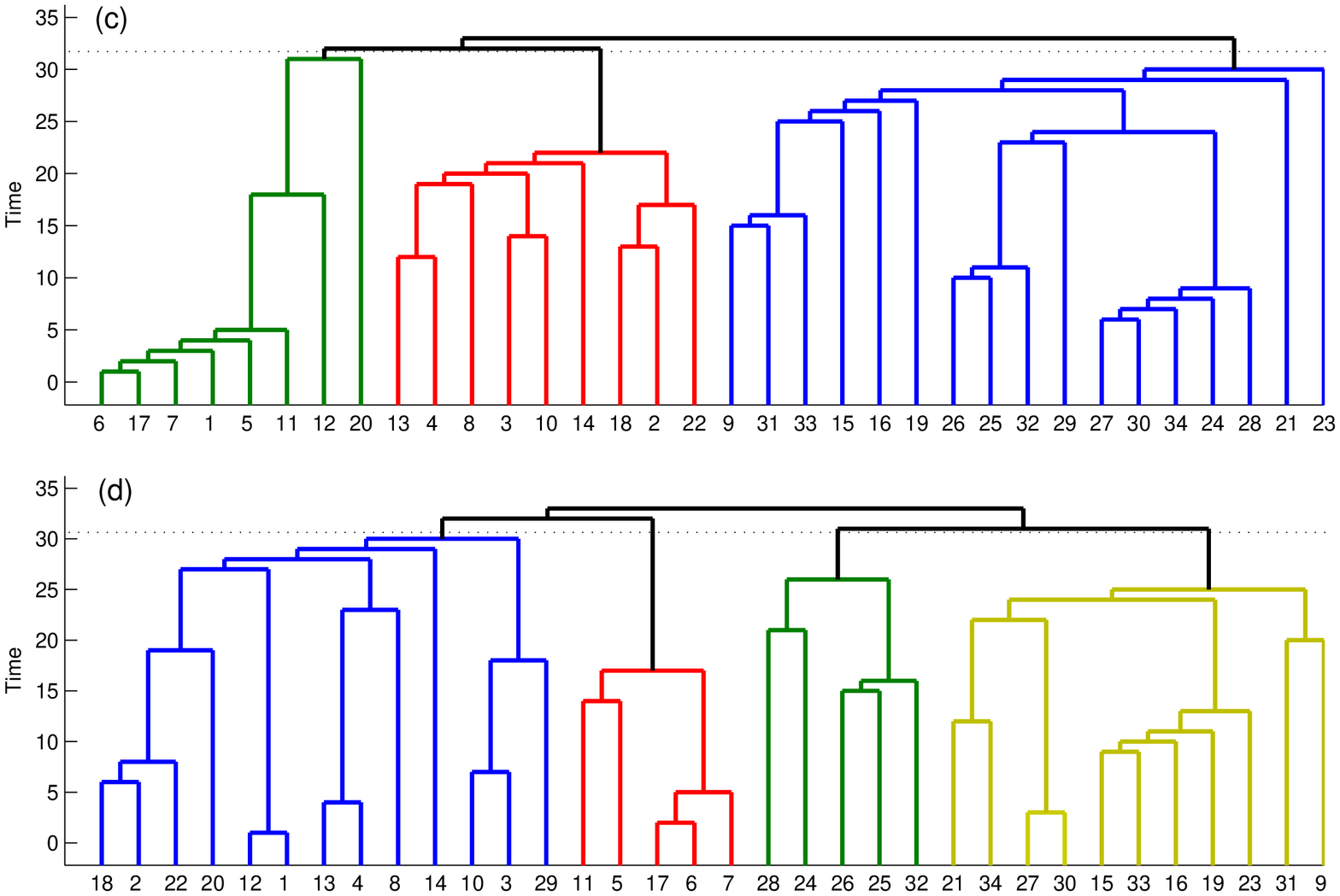}}
 \caption{(Colour online)
(a) Zachary's karate club network (b) Modularity as algorithms
progress (c) Dendrogram representing the progress of fast algorithm,
where formation of large clusters is favoured early (d) Dendrogram
representing the progress of our modification, all clusters are
treated on an equal footing and individual nodes are absorbed into
clusters early.}
 \label{fig:karate}
\end{figure}

Let us analyse carefully how the algorithm proceeds when applied on
the well studied karate club friendship network of Zachary
\cite{Zachary77}. Data on the network was collected over a two year
period before the club split due to an internal dispute during which
some of the members started their own club. The fissure is apparent
in the topology of the network before the split (see Figure
\ref{fig:karate}a), and this data set has become somewhat of a
standard case study for community detection algorithms in the
literature
\cite{Bagrow04,Duch05,Fortunato04,GN,Newman04,Wu03,Zhou03_1,Zhou04,Gfeller05}.

Figure \ref{fig:karate}c shows the dendrogram as generated by the
fast algorithm, with the different colours depicting the partition
at the highest value of $Q=0.3807$. In the first step of the
algorithm, $a_i$ is simply the degree of node $i$ and $e_{ij}$ is 1
for any neighbour pair. Hence, the pair of nodes that will be joined
first is the neighbour pair that has the smallest product of
degrees. In the case of the karate club network, these are nodes $6$
and $17$ with degrees $3$ and $2$ respectively. Note that once a
community has joined with another, the resulting community tends to
join again, since the first term of \ref{dqpure}, $e_{ij}$, tends to
be increased by the joining of neighbouring communities, especially
in networks with high clustering. So, the cluster of nodes $6$ and
$17$ absorb their common neighbour, node $7$. This larger cluster
now has an even larger $e_{ij}$ to common neighbours and in the
following steps absorbs nodes 1, 5 and 11, until no common
neighbours exist. This process occurs in a similar fashion for nodes
24, 27, 28, 30 and 34. We observe that when choosing the pair of
communities to be joined, large communities are favoured at the
expense of smaller ones. In turn, this leads to the formation of a
few large clusters in networks where a larger number of smaller
clusters may represent the real community structure better.

To avoid this and to treat each community as equal, we normalise
$dQ$ by the number of links:

\begin{equation}
dQ'_{ij}=\frac{dQ_{ij}}{a_i}=\frac{2}{a_i}(e_{ij}-\frac{a_ia_j}{2l_{total}})
\label{dqpure}
\end{equation}

It is important to note that while the pair of nodes with the
largest value of $dQ'$ is chosen, the real value of $Q$ must be
calculated at each step using the original $dQ$, or measuring the
value of $Q$ explicitly. Note that as opposed to the original
formulation, this measure is asymmetric, that is $dQ'_{ij}\neq
dQ'_{ji}$. But, the implementation of the algorithm ensures that
both $dQ'_{ij}$ and $dQ'_{ji}$ are considered when choosing the pair
of communities to join, and, since we are interested in only the
largest value of $dQ'$ at each step, this poses no problem. In
essence, the modified algorithm is able to take a different path in
the partition space from the original, in part due to this
asymmetry. For each possible merging of neighbouring communities,
there exists only one value of $dQ$, whereas $dQ'$ takes two
different values, if the two communities have a different number of
links $a_i\neq a_j$.

This normalisation insures that clusters with fewer links have the
largest values of $dQ'$, and therefore are joined earlier. Taking
the karate club network as an example again, we see that
neighbouring nodes where one neighbour has the smallest degree are
joined first. This ensures that nodes with only one link are joined
at the beginning of the process, such as node 12 (see Fig.
\ref{fig:karate}d.). Curiously, using another method based on
synchronisation recently proposed by two of us produces a very
similar dendrogram \cite{Arenas05}. We argue that this is a better
way to proceed. A partition containing a single node will always
contribute negatively to the value of $Q$, even if the degree of
that node is 1. For example in \cite{Donetti04} the authors find a
partition with $Q=0.412$ which has node 12 as a separate community,
using an entirely different method for exploring the partition
space. But, $Q_{i=12}=-1/78$ and the same partition, only with node
12 contained within it's neighbour community, has $Q=0.418$
\footnote{Such a partition is found by a more exhaustive search of
the partition space, by using for example the EO algorithm by Duch
and Arenas \cite{Duch05}.}.

While the NF algorithm also ensures that single node partitions are
not found in the optimal state, our modification performs this
absorption much earlier. This means that in the first few steps of
our algorithm will inevitably appear to performing worse than the NF
algorithm. As it progresses, however, it overtakes the NF algorithm
in terms of $Q$, as we can see in Figure \ref{fig:karate}b. Indeed,
we find that when our modification does not match the performance of
the NF algorithm in terms of $Q$, it improves it.

\section{Testing the modification}
\label{sec:results}

To test the performance of the modification proposed, we have
applied the algorithm on several networks, both ad-hoc and real. To
begin with we look at networks with four equal sized communities, as
described in \cite{NG}.

As $z_{out}/k$ increases, the modularity of the pre-defined
partition decreases as $Q=3/4-z_{out}/k$ irrespective of network or
community size. Figure \ref{fig:4coms}a shows the expected
modularity value compared with those  found by the NF algorithm and
our modification. For low values of $z_{out}/k$ both algorithms find
communities with the value of $Q$ following the expected value
closely. For higher $z_{out}/k$ these values deviate from the
expected value as the communities found by the algorithm do not
correspond exactly to pre-defined communities. In fact, as
$z_{out}/k$ increases above 0.5 the pre-defined partitions give a
lower value of $Q$ than those found by the algorithm, which tend
towards the value that random networks exhibit due to fluctuations
\cite{Guimera04}. The values of $Q$ found by our modification is
very similar to those found by the NF algorithm.

The deviation between pre-defined and found partitions is seen more
clearly by looking at the mutual information measure $I(A,B)$ in the
lower part of Figure \ref{fig:4coms}. As $z_{out}/k$ increases
beyond the point where communities are well defined, the amount of
information about community structure the algorithms are able to
extract decreases. When the communities found have hardly any
relation to pre-defined ones, as is the case for high $z_{out}/k$,
$I(A,B)$ tends to zero. As network size increases however, the
algorithms are able to extract less information from the network
structure. This supports the suggestion that communities in these
networks become more fuzzy as their size increases. Once again, our
modification performs very similarly to the NF algorithm.

It seems logical that both algorithms perform with similar accuracy
for these networks. As we have seen in \ref{sec:modfast} the NF
algorithm seems to favour the formation of larger communities.
However when the communities to be found are all of the same size,
one would expect it to perform quite well. Our modification has
little effect in this case.

\begin{figure}
\centerline{
\includegraphics*[width=0.5\columnwidth]{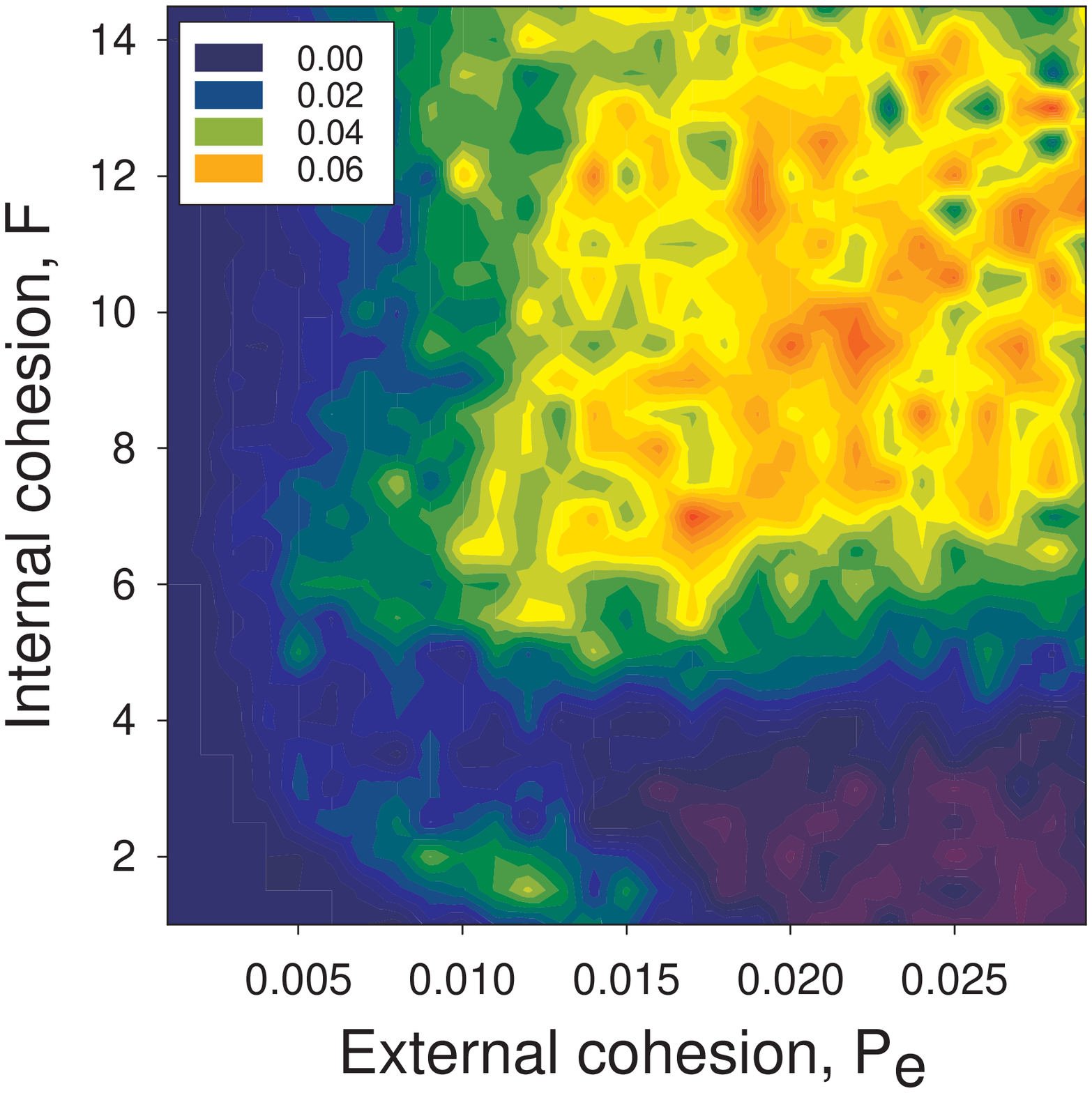}
\includegraphics*[width=0.5\columnwidth]{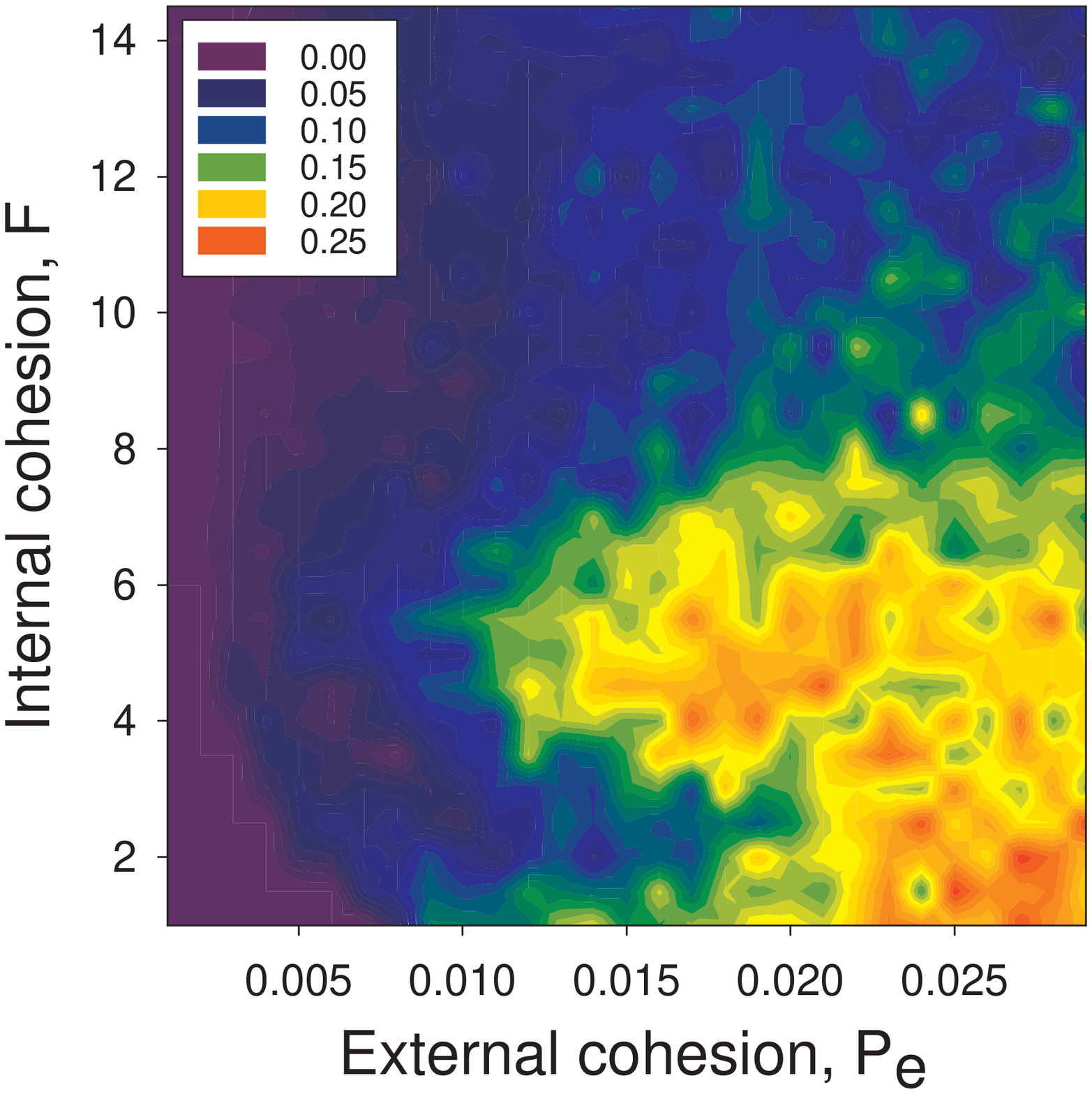}
}\centerline{(a)\hspace{0.5\columnwidth}(b)}
\caption{
    (Colour online) Difference in performance between the NF
    algorithm and our modification (a) proportion of improvement in Q (b)
    proportion of improvement in I(A,B). Our modified algorithm outperforms the
    NF algorithm in all parts of parameter space, but the difference is
    most pronounced for low values of $P_e$ and high values $F$, i.e.
    where the communities are fuzzy.}
    \label{fig:resadhocdiff}
\end{figure}

The difference between the algorithms appears when communities of
different sizes are present within the network. Using the network
construction method proposed in Section \ref{sec:bench}, we study
the performance of the algorithm on networks with 21 communities.
The communities are chosen by hand, with one community of 128 nodes,
four communities with 32 nodes each and 16 communities containing 8
nodes each. This corresponds to a size distribution which follows a
power law (with only three points), where the exponent is -1. In
Figure \ref{fig:resadhocdiff} we show the difference in performance
between the NF algorithm and our modification. They are compared
both in terms of modularity and mutual information. Our modification
performs better in all parts of the parameter space, with some
regions showing up to $25 \%$ improvement over the original
algorithm. The regions where the improvement is largest are those
where the communities are fuzzy, that is, for high values of
external cohesion $P_{s}$ and low values of internal cohesion $F$.

This suggests that our modified algorithm will perform better in
real networks, where the size of communities is highly heterogeneous
and the community structure is fuzzy. To check this, we also
performed tests on some real networks. Table \ref{tab:compare} shows
the comparison of our modified algorithms with Newman's original
formulation and, where possible with the extremal optimisation
algorithm. We looked at the network of Jazz bands with nodes
representing the bands, and links between bands representing at
least one musician that played in both \cite{Gleiser03}; the e-mail
network of University Rovira i Virgili \cite{Guimera03b} where
e-mail addresses are connected by exchanging messages; and the
network of users of the pretty good privacy (PGP) algorithm for
secure information transactions \cite{Guardiola02}. These are medium
sized networks and are still tractable with the Extremal
Optimisation (EO) algorithm \cite{Duch05}, which has a larger
running time scaling as $O(n^2\log{n})$. In these networks, the EO
algorithm clearly performs best out of the three, which is no
surprise since it explores much more of the partition space than
either of the others. It is, however, impractical to use in very
large networks due to running time. In large networks such as the
co-authorship network of the arXiv preprint database
\cite{Newman01a}, the network of web pages within the nd.edu domain
\cite{Albert99}, or the actor network \cite{Barabasi99}, our
algorithm is still able to run in a reasonable time. It improves on
the results of the NF algorithm, finding partitions up to $16\%$
better in terms of $Q$, with no tradeoff in speed.

\begin{table}
 \caption{Table of optimal modularity values
  obtained by the Extremal Optimisation algorithm, $Q_{EO}$ \cite{Duch05},
  the NF algorithm, $Q_{NF}$ \cite{Newman04a}, and the modification
  presented here, $Q_{QM}$.}
 \centering\begin{tabular}{|c||c|c|c|c|} \hline
Network & Size & $Q_{EO}$ & $Q_{NF}$ & $Q_M$ \\
 \hline\hline
    Zachary & 34 & 0.4188 & 0.381 & 0.4087 \\
    Jazz bands  & 198 & 0.4452 & 0.4389 & 0.4409  \\
    E-mail & 1144 & 0.5738 & 0.4796 & 0.5569 \\
    PGP & 10680 & 0.8459 & 0.7329 & 0.7462 \\
    arXiv & 44337 & N/A & 0.7165 & 0.7606 \\
    WWW & 325729 & N/A & 0.9269 & 0.9403 \\
    Actor & 374511 & N/A & 0.6829 & 0.7194 \\
 \hline
  \end{tabular}
  \label{tab:compare}
\end{table}

\section{Conclusion}

To conclude, in this paper we have proposed a more realistic
benchmark test for community detection algorithms in complex
networks which takes into account the heterogeneity of community
size observed in real networks. We have also shown that Newman's
fast community detection algorithm tends to favour the creation of
large communities at the expense of smaller ones. We propose a
simple modification of the fast algorithm which can ensure that
communities of differing sizes are treated on an equal footing, thus
side-stepping this potential problem. Upon comparing the sensitivity
of our modification to that of the original algorithm, we saw that
they perform almost identically in ad-hoc networks with communities
of equal size. However, when compared using the proposed benchmark
test, the improvement in sensitivity increases. Therefore, we claim
that the heterogeneity in community size should be considered when
evaluating community detection algorithms.

Furthermore, we have seen that our modified algorithm gives improved
results for all real networks studied. This improvement is up to
16\% in some studied networks. The improvement in results comes at
no extra computational cost, and a reasonable implementation of the
algorithm will run in $O(n\log^2n)$ time. We recommend the use of
this simple modification for the study of community structure in
very large complex networks.

\section{Acknowledgements}
We thank J. Duch and M. Bogunya for useful discussions and Mark
Newman for providing the arXiv network data. This work has been
supported by DGES of the Spanish Government Grant No. BFM-2003-08258
and EC-FET Open Project No. IST-2001-33555. L.D. gratefully
acknowledges funding from Generalitat de Catalunya.


\begin{thebibliography}{41}
\expandafter\ifx\csname
natexlab\endcsname\relax\def\natexlab#1{#1}\fi
\expandafter\ifx\csname bibnamefont\endcsname\relax
  \def\bibnamefont#1{#1}\fi
\expandafter\ifx\csname bibfnamefont\endcsname\relax
  \def\bibfnamefont#1{#1}\fi
\expandafter\ifx\csname citenamefont\endcsname\relax
  \def\citenamefont#1{#1}\fi
\expandafter\ifx\csname url\endcsname\relax
  \def\url#1{\texttt{#1}}\fi
\expandafter\ifx\csname urlprefix\endcsname\relax\def\urlprefix{URL
}\fi \providecommand{\bibinfo}[2]{#2}
\providecommand{\eprint}[2][]{\url{#2}}

\bibitem[{\citenamefont{Barabasi and Albert}(2002)}]{BARev}
\bibinfo{author}{\bibfnamefont{A.~L.} \bibnamefont{Barabasi}} \bibnamefont{and}
  \bibinfo{author}{\bibfnamefont{R.}~\bibnamefont{Albert}},
  \bibinfo{journal}{Review of Modern Physics} \textbf{\bibinfo{volume}{74}},
  \bibinfo{pages}{47} (\bibinfo{year}{2002}).

\bibitem[{\citenamefont{Dorogovtsev and Mendes}(2003)}]{DMRev}
\bibinfo{author}{\bibfnamefont{S.~N.} \bibnamefont{Dorogovtsev}}
  \bibnamefont{and} \bibinfo{author}{\bibfnamefont{J.~F.~F.}
  \bibnamefont{Mendes}}, \emph{\bibinfo{title}{Evolution of Networks: From
  biological nets to the internet and WWW}} (\bibinfo{publisher}{Oxford
  University Press, Oxford}, \bibinfo{year}{2003}).

\bibitem[{\citenamefont{Newman}(2003)}]{NRev}
\bibinfo{author}{\bibfnamefont{M.~E.~J.} \bibnamefont{Newman}},
  \bibinfo{journal}{SIAM Review} \textbf{\bibinfo{volume}{45}},
  \bibinfo{pages}{167} (\bibinfo{year}{2003}).

\bibitem[{\citenamefont{Strogatz}(2001)}]{Strogatz01}
\bibinfo{author}{\bibfnamefont{S.~H.} \bibnamefont{Strogatz}},
  \bibinfo{journal}{Nature} \textbf{\bibinfo{volume}{410}},
  \bibinfo{pages}{268} (\bibinfo{year}{2001}).

\bibitem[{\citenamefont{Boccaletti et~al.}(2006)\citenamefont{Boccaletti,
  Latora, Moreno, Chavez, and Hwang}}]{Boccaletti06}
\bibinfo{author}{\bibfnamefont{S.}~\bibnamefont{Boccaletti}},
  \bibinfo{author}{\bibfnamefont{V.}~\bibnamefont{Latora}},
  \bibinfo{author}{\bibfnamefont{Y.}~\bibnamefont{Moreno}},
  \bibinfo{author}{\bibfnamefont{M.}~\bibnamefont{Chavez}}, \bibnamefont{and}
  \bibinfo{author}{\bibfnamefont{D.-U.} \bibnamefont{Hwang}},
  \bibinfo{journal}{Physics Reports} p. \bibinfo{pages}{in press}
  (\bibinfo{year}{2006}).

\bibitem[{\citenamefont{Boss et~al.}(2003)\citenamefont{Boss, Elsinger, Summer,
  and Thurner}}]{Thurner04}
\bibinfo{author}{\bibfnamefont{M.}~\bibnamefont{Boss}},
  \bibinfo{author}{\bibfnamefont{H.}~\bibnamefont{Elsinger}},
  \bibinfo{author}{\bibfnamefont{M.}~\bibnamefont{Summer}}, \bibnamefont{and}
  \bibinfo{author}{\bibfnamefont{S.}~\bibnamefont{Thurner}},
  \bibinfo{journal}{cond-mat/0309582}  (\bibinfo{year}{2003}).

\bibitem[{\citenamefont{Ravasz et~al.}(2002)\citenamefont{Ravasz, Somera,
  Mongru, Oltvai, and Barabási}}]{Ravasz02}
\bibinfo{author}{\bibfnamefont{E.}~\bibnamefont{Ravasz}},
  \bibinfo{author}{\bibfnamefont{A.~L.} \bibnamefont{Somera}},
  \bibinfo{author}{\bibfnamefont{D.~A.} \bibnamefont{Mongru}},
  \bibinfo{author}{\bibfnamefont{Z.~N.} \bibnamefont{Oltvai}},
  \bibnamefont{and} \bibinfo{author}{\bibfnamefont{A.-L.}
  \bibnamefont{Barabási}}, \bibinfo{journal}{Science}
  \textbf{\bibinfo{volume}{297}}, \bibinfo{pages}{1551} (\bibinfo{year}{2002}).

\bibitem[{\citenamefont{Guimer\`{a} and
  Amaral}(2005{\natexlab{a}})}]{Guimera05a}
\bibinfo{author}{\bibfnamefont{R.}~\bibnamefont{Guimer\`{a}}} \bibnamefont{and}
  \bibinfo{author}{\bibfnamefont{L.~N.~A.} \bibnamefont{Amaral}},
  \bibinfo{journal}{Nature} \textbf{\bibinfo{volume}{433}},
  \bibinfo{pages}{895} (\bibinfo{year}{2005}{\natexlab{a}}).

\bibitem[{\citenamefont{Guimer\`{a} and
  Amaral}(2005{\natexlab{b}})}]{Guimera05b}
\bibinfo{author}{\bibfnamefont{R.}~\bibnamefont{Guimer\`{a}}} \bibnamefont{and}
  \bibinfo{author}{\bibfnamefont{L.~N.~A.} \bibnamefont{Amaral}},
  \bibinfo{journal}{JSTAT} p. \bibinfo{pages}{P02001}
  (\bibinfo{year}{2005}{\natexlab{b}}).

\bibitem[{\citenamefont{Girvan and Newman}(2002)}]{GN}
\bibinfo{author}{\bibfnamefont{M.}~\bibnamefont{Girvan}} \bibnamefont{and}
  \bibinfo{author}{\bibfnamefont{M.~E.~J.} \bibnamefont{Newman}},
  \bibinfo{journal}{Publications of the National Academy of Sciences USA}
  \textbf{\bibinfo{volume}{99}}, \bibinfo{pages}{7821} (\bibinfo{year}{2002}).

\bibitem[{\citenamefont{Kernighan and Lin}(1970)}]{KernighanLin}
\bibinfo{author}{\bibfnamefont{B.~W.} \bibnamefont{Kernighan}}
  \bibnamefont{and} \bibinfo{author}{\bibfnamefont{S.}~\bibnamefont{Lin}},
  \bibinfo{journal}{The Bell System Tech J} \textbf{\bibinfo{volume}{49}},
  \bibinfo{pages}{291} (\bibinfo{year}{1970}).

\bibitem[{\citenamefont{Pothen}(1996)}]{PothenRev}
\bibinfo{author}{\bibfnamefont{A.}~\bibnamefont{Pothen}},
  \emph{\bibinfo{title}{Parallel Numerical Algorithms}}
  (\bibinfo{publisher}{Kluwer Academic Press}, \bibinfo{year}{1996}), chap.
  \bibinfo{chapter}{Graph partitioning algorithms with applications to
  scientific computing.}

\bibitem[{\citenamefont{Newman}(2004{\natexlab{a}})}]{Newman04}
\bibinfo{author}{\bibfnamefont{M.~E.~J.} \bibnamefont{Newman}},
  \bibinfo{journal}{European Physics Journal B} \textbf{\bibinfo{volume}{38}},
  \bibinfo{pages}{321} (\bibinfo{year}{2004}{\natexlab{a}}).

\bibitem[{\citenamefont{Danon et~al.}(2005)\citenamefont{Danon,
  D\'{i}az-Guilera, Duch, and Arenas}}]{Danon05}
\bibinfo{author}{\bibfnamefont{L.}~\bibnamefont{Danon}},
  \bibinfo{author}{\bibfnamefont{A.}~\bibnamefont{D\'{i}az-Guilera}},
  \bibinfo{author}{\bibfnamefont{J.}~\bibnamefont{Duch}}, \bibnamefont{and}
  \bibinfo{author}{\bibfnamefont{A.}~\bibnamefont{Arenas}},
  \bibinfo{journal}{J. Stat. Mech} p. \bibinfo{pages}{P09008}
  (\bibinfo{year}{2005}).

\bibitem[{\citenamefont{Reichardt and Bornholdt}(2004)}]{Reichardt04}
\bibinfo{author}{\bibfnamefont{J.}~\bibnamefont{Reichardt}} \bibnamefont{and}
  \bibinfo{author}{\bibfnamefont{S.}~\bibnamefont{Bornholdt}},
  \bibinfo{journal}{Phys. Rev. Lett.} \textbf{\bibinfo{volume}{93}},
  \bibinfo{pages}{218701} (\bibinfo{year}{2004}).

\bibitem[{\citenamefont{Guimer\`{a} et~al.}(2004)\citenamefont{Guimer\`{a},
  Sales, and Amaral}}]{Guimera04}
\bibinfo{author}{\bibfnamefont{R.}~\bibnamefont{Guimera}},
  \bibinfo{author}{\bibfnamefont{M.}~\bibnamefont{Sales-Pardo}}, \bibnamefont{and}
  \bibinfo{author}{\bibfnamefont{L.~N.~A.} \bibnamefont{Amaral}},
  \bibinfo{journal}{Physical Review E} \textbf{\bibinfo{volume}{70}},
  \bibinfo{pages}{025101(R)} (\bibinfo{year}{2004}).

\bibitem[{\citenamefont{Newman}(2004{\natexlab{b}})}]{Newman04a}
\bibinfo{author}{\bibfnamefont{M.~E.~J.} \bibnamefont{Newman}},
  \bibinfo{journal}{Physical Review E} \textbf{\bibinfo{volume}{69}},
  \bibinfo{pages}{066133} (\bibinfo{year}{2004}{\natexlab{b}}).

\bibitem[{\citenamefont{Clauset et~al.}(2004)\citenamefont{Clauset, Newman, and
  Moore}}]{Clauset04}
\bibinfo{author}{\bibfnamefont{A.}~\bibnamefont{Clauset}},
  \bibinfo{author}{\bibfnamefont{M.~E.~J.} \bibnamefont{Newman}},
  \bibnamefont{and} \bibinfo{author}{\bibfnamefont{C.}~\bibnamefont{Moore}},
  \bibinfo{journal}{Physical Review E} \textbf{\bibinfo{volume}{70}},
  \bibinfo{pages}{066111} (\bibinfo{year}{2004}).

\bibitem[{\citenamefont{Bagrow and Bollt}(2005)}]{Bagrow04}
\bibinfo{author}{\bibfnamefont{J.~P.} \bibnamefont{Bagrow}} \bibnamefont{and}
  \bibinfo{author}{\bibfnamefont{E.~M.} \bibnamefont{Bollt}},
  \bibinfo{journal}{Physical Review E} \textbf{\bibinfo{volume}{72}},
  \bibinfo{pages}{046108} (\bibinfo{year}{2005}).

\bibitem[{\citenamefont{Duch and Arenas}(2005)}]{Duch05}
\bibinfo{author}{\bibfnamefont{J.}~\bibnamefont{Duch}} \bibnamefont{and}
  \bibinfo{author}{\bibfnamefont{A.}~\bibnamefont{Arenas}},
  \bibinfo{journal}{Physical Review E} \textbf{\bibinfo{volume}{72}},
  \bibinfo{pages}{027104} (\bibinfo{year}{2005}).

\bibitem[{\citenamefont{Wu and Huberman}(2004)}]{Wu03}
\bibinfo{author}{\bibfnamefont{F.}~\bibnamefont{Wu}} \bibnamefont{and}
  \bibinfo{author}{\bibfnamefont{B.}~\bibnamefont{Huberman}},
  \bibinfo{journal}{Eurpean Physics Journal B} \textbf{\bibinfo{volume}{38}},
  \bibinfo{pages}{331} (\bibinfo{year}{2004}).

\bibitem[{\citenamefont{Newman and Girvan}(2004)}]{NG}
\bibinfo{author}{\bibfnamefont{M.~E.~J.} \bibnamefont{Newman}}
  \bibnamefont{and} \bibinfo{author}{\bibfnamefont{M.}~\bibnamefont{Girvan}},
  \bibinfo{journal}{Physical Review E} \textbf{\bibinfo{volume}{69}},
  \bibinfo{pages}{026113} (\bibinfo{year}{2004}).

\bibitem[{\citenamefont{Guimer\'{a} et~al.}(2003)\citenamefont{Guimer\'{a},
  Danon, Diaz-Guilera, Giralt, and Arenas}}]{Guimera03b}
\bibinfo{author}{\bibfnamefont{R.}~\bibnamefont{Guimera}},
  \bibinfo{author}{\bibfnamefont{L.}~\bibnamefont{Danon}},
  \bibinfo{author}{\bibfnamefont{A.}~\bibnamefont{Diaz-Guilera}},
  \bibinfo{author}{\bibfnamefont{F.}~\bibnamefont{Giralt}}, \bibnamefont{and}
  \bibinfo{author}{\bibfnamefont{A.}~\bibnamefont{Arenas}},
  \bibinfo{journal}{Physical Review E} \textbf{\bibinfo{volume}{68}}
  (\bibinfo{year}{2003}).

\bibitem[{\citenamefont{Palla et~al.}(2005)\citenamefont{Palla, Derenyi,
  Farkas, and Vicsek}}]{Palla05}
\bibinfo{author}{\bibfnamefont{G.}~\bibnamefont{Palla}},
  \bibinfo{author}{\bibfnamefont{I.}~\bibnamefont{Derenyi}},
  \bibinfo{author}{\bibfnamefont{I.}~\bibnamefont{Farkas}}, \bibnamefont{and}
  \bibinfo{author}{\bibfnamefont{T.}~\bibnamefont{Vicsek}},
  \bibinfo{journal}{Nature} \textbf{\bibinfo{volume}{435}},
  \bibinfo{pages}{814} (\bibinfo{year}{2005}).

\bibitem[{\citenamefont{Arenas et~al.}(2004)\citenamefont{Arenas, Danon,
  Diaz-Guilera, Gleiser, and Guimer\`a}}]{Arenas03}
\bibinfo{author}{\bibfnamefont{A.}~\bibnamefont{Arenas}},
  \bibinfo{author}{\bibfnamefont{L.}~\bibnamefont{Danon}},
  \bibinfo{author}{\bibfnamefont{A.}~\bibnamefont{Diaz-Guilera}},
  \bibinfo{author}{\bibfnamefont{P.~M.} \bibnamefont{Gleiser}},
  \bibnamefont{and}
  \bibinfo{author}{\bibfnamefont{R.}~\bibnamefont{Guimer\`a}},
  \bibinfo{journal}{European Physical Journal B} \textbf{\bibinfo{volume}{38}},
  \bibinfo{pages}{373} (\bibinfo{year}{2004}).

\bibitem[{\citenamefont{Gleiser and Danon}(2003)}]{Gleiser03}
\bibinfo{author}{\bibfnamefont{P.}~\bibnamefont{Gleiser}} \bibnamefont{and}
  \bibinfo{author}{\bibfnamefont{L.}~\bibnamefont{Danon}},
  \bibinfo{journal}{Advances in Complex Systems} \textbf{\bibinfo{volume}{6}},
  \bibinfo{pages}{565} (\bibinfo{year}{2003}).

\bibitem[{\citenamefont{Fred and Jain}(2003)}]{Fred03}
\bibinfo{author}{\bibfnamefont{A.~L.~N.} \bibnamefont{Fred}} \bibnamefont{and}
  \bibinfo{author}{\bibfnamefont{A.~K.} \bibnamefont{Jain}},
  \bibinfo{journal}{Proc. IEEE Computer Society Conference on Computer Vision
  and Pattern Recognition, CVPR, USA} pp. \bibinfo{pages}{II--128--133}
  (\bibinfo{year}{2003}).

\bibitem[{\citenamefont{Kuncheva and Hadjitodorov}(2004)}]{Kuncheva04}
\bibinfo{author}{\bibfnamefont{L.~I.} \bibnamefont{Kuncheva}} \bibnamefont{and}
  \bibinfo{author}{\bibfnamefont{S.~T.} \bibnamefont{Hadjitodorov}},
  \bibinfo{journal}{Systems, Man and Cybernetics, 2004 IEEE International
  Conference} \textbf{\bibinfo{volume}{2}}, \bibinfo{pages}{1214}
  (\bibinfo{year}{2004}).

\bibitem[{\citenamefont{Everitt}(1993)}]{Everitt93}
\bibinfo{author}{\bibfnamefont{B.}~\bibnamefont{Everitt}},
  \emph{\bibinfo{title}{Cluster Analysis, 3rd edition}}
  (\bibinfo{publisher}{Edward Arnold, London}, \bibinfo{year}{1993}).

\bibitem[{\citenamefont{Scott}(2000)}]{Scott00}
\bibinfo{author}{\bibfnamefont{J.}~\bibnamefont{Scott}},
  \emph{\bibinfo{title}{Social Network Analysis, a handboook}}
  (\bibinfo{publisher}{SAGE publications, London, 2nd edition},
  \bibinfo{year}{2000}).

\bibitem[{\citenamefont{Zachary}(1977)}]{Zachary77}
\bibinfo{author}{\bibfnamefont{W.~W.} \bibnamefont{Zachary}},
  \bibinfo{journal}{Journal of Anthropological Research}
  \textbf{\bibinfo{volume}{33}}, \bibinfo{pages}{452} (\bibinfo{year}{1977}).

\bibitem[{\citenamefont{Fortunato et~al.}(2004)\citenamefont{Fortunato, Latora,
  and Marchiori}}]{Fortunato04}
\bibinfo{author}{\bibfnamefont{S.}~\bibnamefont{Fortunato}},
  \bibinfo{author}{\bibfnamefont{V.}~\bibnamefont{Latora}}, \bibnamefont{and}
  \bibinfo{author}{\bibfnamefont{M.}~\bibnamefont{Marchiori}},
  \bibinfo{journal}{Physical Review E} \textbf{\bibinfo{volume}{70}}
  (\bibinfo{year}{2004}).

\bibitem[{\citenamefont{Zhou}(2003)}]{Zhou03_1}
\bibinfo{author}{\bibfnamefont{H.}~\bibnamefont{Zhou}},
  \bibinfo{journal}{Physical Review E} \textbf{\bibinfo{volume}{67}},
  \bibinfo{pages}{041908} (\bibinfo{year}{2003}).

\bibitem[{\citenamefont{Zhou and Lipowsky}(2004)}]{Zhou04}
\bibinfo{author}{\bibfnamefont{H.}~\bibnamefont{Zhou}} \bibnamefont{and}
  \bibinfo{author}{\bibfnamefont{R.}~\bibnamefont{Lipowsky}},
  \bibinfo{journal}{Lecture Notes in Computer Sciences}
  \textbf{\bibinfo{volume}{3038}}, \bibinfo{pages}{1062 }
  (\bibinfo{year}{2004}).

\bibitem[{\citenamefont{Gfeller et~al.}(2005)\citenamefont{Gfeller, Chappelier,
  and De~Los~Rios}}]{Gfeller05}
\bibinfo{author}{\bibfnamefont{D.}~\bibnamefont{Gfeller}},
  \bibinfo{author}{\bibfnamefont{J.-C.} \bibnamefont{Chappelier}},
  \bibnamefont{and}
  \bibinfo{author}{\bibfnamefont{P.}~\bibnamefont{DeLosRios}},
  \bibinfo{journal}{Physical Review E} \textbf{\bibinfo{volume}{72}},
  \bibinfo{pages}{056135} (\bibinfo{year}{2005}).

\bibitem[{\citenamefont{Arenas et~al.}(2005)\citenamefont{Arenas, Diaz-Guilera,
  and Perez-Vicente}}]{Arenas05}
\bibinfo{author}{\bibfnamefont{A.}~\bibnamefont{Arenas}},
  \bibinfo{author}{\bibfnamefont{A.}~\bibnamefont{Diaz-Guilera}},
  \bibnamefont{and} \bibinfo{author}{\bibfnamefont{C.~J.}
  \bibnamefont{Perez-Vicente}}, pp. \bibinfo{pages}{cond--mat/0511730}
  (\bibinfo{year}{2005}).

\bibitem[{\citenamefont{Donetti and \protect{Mu\~{n}oz}}(2004)}]{Donetti04}
\bibinfo{author}{\bibfnamefont{L.}~\bibnamefont{Donetti}} \bibnamefont{and}
  \bibinfo{author}{\bibfnamefont{M.~A.} \bibnamefont{\protect{Mu\~{n}oz}}},
  \bibinfo{journal}{Journal of Statistical Mechanics: Theory and Experiment}
  (\bibinfo{year}{2004}).

\bibitem[{\citenamefont{Guardiola et~al.}(2002)\citenamefont{Guardiola,
  Guimer\`{a}, Arenas, Diaz-Guilera, Streib, and Amaral}}]{Guardiola02}
\bibinfo{author}{\bibfnamefont{X.}~\bibnamefont{Guardiola}},
  \bibinfo{author}{\bibfnamefont{R.}~\bibnamefont{Guimer\`{a}}},
  \bibinfo{author}{\bibfnamefont{A.}~\bibnamefont{Arenas}},
  \bibinfo{author}{\bibfnamefont{A.}~\bibnamefont{Diaz-Guilera}},
  \bibinfo{author}{\bibfnamefont{D.}~\bibnamefont{Streib}}, \bibnamefont{and}
  \bibinfo{author}{\bibfnamefont{L.}~\bibnamefont{Amaral}},
  \bibinfo{journal}{cond-mat/0206240}  (\bibinfo{year}{2002}).

\bibitem[{\citenamefont{Newman}(2001)}]{Newman01a}
\bibinfo{author}{\bibfnamefont{M.~E.~J.} \bibnamefont{Newman}},
  \bibinfo{journal}{Physical Review E} \textbf{\bibinfo{volume}{64}},
  \bibinfo{pages}{016132} (\bibinfo{year}{2001}).

\bibitem[{\citenamefont{Albert et~al.}(1999)\citenamefont{Albert, Jeong, and
  Barab\'{a}si}}]{Albert99}
\bibinfo{author}{\bibfnamefont{R.}~\bibnamefont{Albert}},
  \bibinfo{author}{\bibfnamefont{H.}~\bibnamefont{Jeong}}, \bibnamefont{and}
  \bibinfo{author}{\bibfnamefont{A.-L.} \bibnamefont{Barab\'{a}si}},
  \bibinfo{journal}{Nature} \textbf{\bibinfo{volume}{401}},
  \bibinfo{pages}{130} (\bibinfo{year}{1999}).

\bibitem[{\citenamefont{Barab\'{a}si and Albert}(1999)}]{Barabasi99}
\bibinfo{author}{\bibfnamefont{A.-L.} \bibnamefont{Barab\'{a}si}}
  \bibnamefont{and} \bibinfo{author}{\bibfnamefont{R.}~\bibnamefont{Albert}},
  \bibinfo{journal}{Science} \textbf{\bibinfo{volume}{286}},
  \bibinfo{pages}{509} (\bibinfo{year}{1999}).

\end{thebibliography}
\end{document}